\documentclass[twocolumn]{autart-2}
\usepackage{cite}
\usepackage{amsmath,amssymb,amsfonts}
\usepackage{graphicx}
\usepackage{hyperref}
\usepackage{multirow}
\usepackage{textcomp}
\usepackage{mathbbol}
\usepackage{dirtytalk}     
\usepackage{xcolor}
\usepackage{graphicx}
\usepackage{amsmath}
\usepackage{mathtools}
\usepackage{verbatim}
\usepackage{amssymb}
\usepackage{array}
\usepackage{subcaption}
\usepackage[skip=3pt]{caption}
\newtheorem{lemma}{\textbf{Lemma}}

\newtheorem{corollary}{\textbf{Corollary}}
\newtheorem{definition}{\textbf{Definition}}
\newtheorem{remark}{\textbf{Remark}}
\newtheorem{expm}{\textbf{Example}}
\newenvironment{proof}{{\emph{\textbf{Proof:}} }}{\hfill $\square$}
\DeclareMathOperator\Spec{Spec}

\begin{document}
\begin{frontmatter}
\title{\LARGE \bf
A Signed Friedkin-Johnsen Model for Arbitrary Network Topologies} 
\vspace{-1cm}
\author{Aashi Shrinate}\ead{aashis21@iitk.ac.in}, 
\author{Twinkle Tripathy}\ead{ttripathy@iitk.ac.in} 

\address{Department of Electrical Engineering, Indian Institute of Technology Kanpur, Uttar Pradesh, India 208016.} 
\vspace{-0.8cm}

\begin{abstract}
The paper presents an opposing rule-based signed Friedkin-Johnsen (SFJ) model for the evolution of opinions in arbitrary network topologies with signed interactions and stubborn agents. The primary objective of the paper is to analyse the emergent behaviours of the agents under the proposed rule and to identify the key agents which contribute to the final opinions, characterised as \textit{influential} agents. We start by presenting some convergence results which show how the opinions of the agents evolve for a signed network with any arbitrary topology. Throughout the paper, we classify the agents as opinion leaders (sinks in the associated condensation graph) and followers (the rest). In general, it has been shown in the literature that opinion leaders and stubborn agents drive the opinions of the group. However, the addition of signed interactions reveals interesting behaviours wherein opinion leaders can now become \textit{non-influential} or less \textit{influential}. Further, while the stubborn agents always continue to remain influential, they might become less \textit{influential} owing to signed interactions. Additionally, the signed interactions can drive the opinions of the agents outside of the convex hull of their initial opinions. Thereafter, we propose the \textit{absolute influence centrality measure} which allows us to quantify the overall influence of all the agents in the network and also identify the most influential agents. Unlike most of the existing measures, it is applicable to any network topology and considers the effect of both stubbornness and signed interactions. Finally, simulations are presented for the Bitcoin Alpha dataset to elaborate the proposed results.
\end{abstract}

\end{frontmatter}

\section{INTRODUCTION}
Consider a group of connected individuals communicating with each other. The opinion of an individual undergoes a natural transformation through the interactions within the group. 
The analysis of the resulting behaviours is a complex problem but an important one for our society, especially when social networks are being employed to influence consumer behaviours \cite{voramontri2019impact}, voting preferences \cite{fernandez2014voter} and shaping public opinions via disinformation campaigns \cite{Gorodnichenko2021} among others.
Several models have been proposed to study opinion formation in a network of interacting agents, e.g. averaging based the DeGroot's model \cite{degroot1974reaching}, Friedkin-Johnsen (FJ) model \cite{FRIEDKIN1997209}, homophily-based Hegselmann-Krause model \cite{rainer2002opinion}, \textit{etc}. However, the FJ model, an extension of DeGroot's model, is popular due to its analytical tractability and its ability to accurately predict individuals' opinions in human-subject experiments.

The FJ model accounts for disagreement, the most commonly observed behaviour in a society, by introducing agent(s) who are stubborn in their prejudices. 
Another peculiar aspect of such an opinion formation process is that the opinion value at which the convergence occurs often depends only on the initial states of certain \textit{influential} agents in the network. 
The authors in \cite{Community_Cleavage} proposed the notion of \textit{influence centrality} (IC) to quantify the contribution of these influential agents in the final opinion in the FJ framework. The authors in \cite{GHADERI20143209,gionis2013opinion} quantify the impact of stubborn agents on the final opinions (IC) based on hitting probabilities of random walks in undirected and directed social networks, respectively.
In the aforementioned works, each agent is assumed to have a path to a stubborn agent, making only the latter influential. In contrast,
the authors in \cite{parsegov2016novel} and \cite{TIAN2018213} present the conditions for convergence of opinions in an extended FJ framework, which includes agents who do not have a path to any stubborn agent. This framework results in a class of non-stubborn agents who can be influential. Moreover, the authors in \cite{TIAN2018213} present the necessary conditions for such agents to become influential.

The works in \cite{FJ_Model,Community_Cleavage,GHADERI20143209,gionis2013opinion,TIAN2018213,parsegov2016novel} determine the final opinions (equivalently, the IC) considering the network has cooperative interactions, which is admissible in various applications. Social networks, however, generally have both cooperative and competitive interactions, which are represented by signed networks. Opinion evolution under signed interactions in DeGroot's continuous and discrete-time frameworks is explored in the literature   \cite{xia2015structural} and \cite{altafini}, respectively. Two kinds of update rules under signed interactions, the opposing rule \cite{altafini,xia2015structural} and the repelling rule \cite{fontan2022multiagent}, are proposed in the literature. 

The opposing and repelling rules in DeGroot's Framework are used to achieve a variety of behaviours such as consensus \cite{fontan2022multiagent}, bipartite consensus \cite{altafini,xia2015structural} and clustering \cite{priya2024desired,shrinate2023desired}. The authors in \cite{razaq2025signed} propose the repelling rule-based signed FJ (SFJ) model and present the sufficient algebraic conditions that the adjacency matrix must satisfy to ensure convergence under both discrete and continuous-time SFJ models.

In contrast to \cite{razaq2025signed}, this paper examines the opinion formation in signed networks using an opposing-rule based SFJ model. We show that the proposed model ensures convergence in any signed network, making it more suitable for both weakly and strongly connected digraphs than the repelling-rule. Due to the signed interactions in the network and the topological positions of the stubborn agents, various collective behaviours emerge in the proposed framework. By analysing how the opinions converge, we also identify the influential agents in all of these diverse scenarios. Thereafter, to identify the most influential node(s) in the network, we propose a new centrality measure, referred to as the \textit{absolute centrality measure}; its key advantage is that it is defined for any arbitrary network structure and it accounts for the effects of signed interactions, stubbornness and the opinion leaders. With this, we now highlight the major contributions of this work:
 
\begin{itemize}
 \item \textit{A generalised framework:} The opinion evolution model proposed in this work generalises the FJ and DeGroot models in \cite{parsegov2016novel} to a signed network. Additionally, it is applicable to any arbitrary topology of the signed network and for any choice of stubborn agents \cite{Tian2022}. 
\item \textit{Emergent behaviours and influential agents:} Through this study, we reveal two types of \textit{influential agents}: opinion leaders (topologically prominent) and stubborn agents (behaviorally prominent). Under the proposed SFJ model, the most common behaviour is disagreement which is similar to a cooperative scenario. Additionally, the stubborn agents are always influential. Interestingly, however, we show that an opinion leader may not be \textit{influential} at all or can be less influential than a follower (nodes constituting non-sink nodes in the associated condensation graph). However, unlike the cooperative case, the opinions can now lie outside of the convex hull of the initial opinions. 
\item \textit{Absolute centrality measure:}  Despite the popularity of FJ models, none of the existing influence centrality measures account for antagonism and stubborn behaviours. The \textit{absolute centrality measure} proposed in this work bridges this gap in the literature since it is derived from the proposed SFJ model. The highlight of this proposed measure is that it is applicable to any arbitrary topology of a given signed network.
\end{itemize}
The paper has been organised as follows: Sec. \ref{Sec2} discusses the relevant notations and preliminaries. The SFJ opinion model and the classification of the agents are presented in Sec. \ref{Sec3}. Sec. \ref{Sec4} analyses the convergence of opinions, and the absolute influence centrality is presented in Sec. \ref{Sec6}. The simulation results are presented in Sec. \ref{sec:Simulation_Results}. Finally, we conclude the paper in Sec. \ref{sec:con} with some insights into the possible future research directions.
\section{Notations \& Preliminaries}
\label{Sec2}
\subsection{Notations}
\label{subsec:NOT}
The vector $\mathbb{1} (\text{ or }\mathbb{0}) \in \mathbb{R}^n$ denotes a column vector with all entries equal to $1$ (or $0$) of appropriate dimensions. For a given matrix $M =[m_{ij}] \in \mathbb{R}^{n \times n}$, let $\tilde{M}=[|m_{ij}|]$, where $|m_{ij}|$ is the absolute value of $(i,j)^{th}$ entry of $M$ for all $i,j=\{1,2,...,n\}$. A diagonal matrix $M \in \mathbb{R}^{n \times n}$ is denoted by $M=diag([m_1,m_2,...,m_n])$. A matrix $M$ with each entry $m_{ij}>0$ (or $m_{ij}\geq 0$) is positive (or non-negative).
The spectrum of a matrix $M$ is denoted by $\Spec(M)$ and its spectral radius is denoted by $\rho(M)=\max\{ |\lambda| : \lambda \in \Spec(M)\}$.
\subsection{Graph Preliminaries}
\label{subsec:GP}
A digraph is defined as $\mathcal{G}=\{\mathcal{V},\mathcal{E}\}$ where $\mathcal{V}=\{1,2,...,$ $n\}$ is the set of nodes representing the $n$ agents in the network, $\mathcal{E} \subseteq \mathcal{V} \times \mathcal{V}$ is the set of edges which indicate interactions in the network. Matrix $A=[a_{ij}] \in \mathbb{R}^{n \times n}$ is the signed weighted adjacency matrix with $a_{ij} \neq 0$ if and only if $(i,j) \in \mathcal{E}$. The entry $a_{ij}$ equals the edge weight of edge $(i,j) \in \mathcal{E}$. The edge $(i,j) \in \mathcal{E}$ implies that $i$ is the in-neighbour of $j$, and $j$ is the out-neighbour of $i$.  A sink is a node in $\mathcal{G}$ without any out-neighbours. 

A path is an ordered sequence of nodes in which every pair of adjacent nodes forms an edge in set $\mathcal{E}$. 
An undirected graph is connected if there exists a path between every pair of nodes.
A digraph is a strongly connected graph if a directed path exists between every pair of nodes in the graph. A digraph is weakly connected if it is not strongly connected, but its undirected version is connected. 
The condensation graph of a graph $\mathcal{G}$ is defined as $C(\mathcal{G})=(\mathcal{V}_c,\mathcal{E}_c)$. Each node $\mathcal{I} \in \mathcal{V}_c$ is a  strongly connected component (SCC) of graph $\mathcal{G}$, and an edge $(\mathcal{I},\mathcal{J})\in \mathcal{E}_c $ exists if and only if an edge $(i,j)\in \mathcal{E}$ exists in graph $\mathcal{G}$ from node $i \in \mathcal{I}$ to a node $j \in \mathcal{J}$.  A sink of the condensation graph is an  SCC in $\mathcal{G}$ that forms a node in the $C(\mathcal{G})$ without any outgoing edges.

A digraph is structurally balanced (SB) if there exists a bipartition of vertices $\mathcal{V} \text{ such that } \mathcal{V}_1 \cap \mathcal{V}_2=\emptyset$ and $\mathcal{V}_1 \cup \mathcal{V}_2 =\mathcal{V}$ 
with positive interaction $a_{ij} \geq 0$ between nodes  $i,j \in \mathcal{V}_q \ (q\in\{1,2\})$ and negative interaction $a_{ij} \leq 0$ if  $i \in \mathcal{V}_{p}$ and $j \in \mathcal{V}_q,p\neq q, (p,q\in\{1,2\})$. A graph with cooperative interactions is also considered to be SB.
Any graph that is not SB is structurally unbalanced (SUB).

\subsection{Matrix Preliminaries}
\label{subsec:MP}
A matrix $M\in \mathbb{R}^{n \times n}$ is row stochastic if $M$ is non-negative and $M\mathbb{1}=\mathbb{1}$. It is row-substochastic if it is non-negative and its row sums are at most one with at least one being strictly less than one.
A matrix $M \in \mathbb{R}^{n \times n}$ is semi-convergent if the $\lim_{k \to \infty}M^k$ exists. It is convergent if $\lim_{k \to \infty}M^k=\mathbb{0}$. A semi-convergent but not convergent matrix has the following spectral properties:
\begin{itemize}
    \item $1$ is a simple or semi-simple eigenvalue,
    \item all the other eigenvalues have magnitude less than $1$.
\end{itemize} 
\begin{lemma}[\cite{gelfand1941normierte}]
\label{lm:1}
For any $M \in \mathbb{C}^{n \times n}$ and any induced norm $\|.\|$, the Gelfand's formula states that the spectral radius $\rho(M)$ of matrix $M$ is  $\rho(M)=\lim_{k \to \infty}\|M^k\|^{\frac{1}{k}}$.
\end{lemma}

\begin{lemma}
\label{lm:2}
The spectral radius of a matrix $M \in \mathbb{R}^{n \times n}$ and $\Tilde{M}=[|m_{ij}|]$, derived from $M$, satisfy $ \rho(M)\leq \rho(\Tilde{M})$.
\end{lemma}
\begin{proof}
    Since the row sum of $M^k$ is strictly less or equal to the row sum of $\tilde{M}^k$, we get $\|M^k\|_{\infty} \leq \| \Tilde{M}^k\|_{\infty}$. This implies that $\lim_{k \to \infty}\|M^k\|_{\infty}^{1/k} \leq \| \lim_{k \to \infty}\Tilde{M}^{k}\|_{\infty}^{1/k}$. Using the Gelfand's formula from Lemma \ref{lm:1}, we get $\rho(M) \leq \rho(\Tilde{M})$.
\end{proof}
\section{Opinion Dynamics}
\label{Sec3}
In a society, there are often individuals or groups of individuals who have high credibility and expertise in certain domains or wield significant power over the beliefs and actions of a broad audience. They can act as opinion leaders of the group who play a major role in opinion formation. Some examples of real-world opinion leaders include socio-political leaders, successful entrepreneurs and highly reputed scholars. 
The advent of social media has led to the emergence of a new kind of opinion leaders, referred to as \textit{influencer} who shapes public opinion \cite{ding2023electoral} and affects consumer behaviour \cite{voramontri2019impact}, \textit{etc}.

It is also important to note that despite the widespread influence of opinion leaders, certain outliers may exist who resist changes in their perception. 
We refer to them as \textit{stubborn} agents. 
Our objective is to study the evolution of opinions and quantify the \textit{influence} that individuals in a network exert on each other in such a heterogeneous setting.

\subsection{Opposing rule-based SFJ model}
\label{subsec:SFJM}
In this paper, we study the evolution of opinions in a network of agents where certain agents are stubborn with respect to their prejudices. Under the FJ model, an agent in a cooperative network takes a convex combination of its neighbours' opinions and its own prejudices to update its opinions.
However, in discussions on issues of importance such as politics \cite{NEAL2020103}, international relations \cite{harary1961structural}, sports \cite{fink2009off}, \textit{etc.}, individuals may have competing interests, which are denoted by signed interactions.

Consider a signed network $\mathcal{G}$ with agents indexed $1$ to $n$. The opinions of agents at the $k^{th}$ instance are given by the vector $\mathbf{x}(k)=[x_1(k),x_2(k),...,x_n(k)] \ \in \mathbb{R}^n$ where $x_i(k)$ is the opinion of the $i^{th}$ agent in the group.  {The opinion of an agent in a group of $n$ heterogeneous agents is governed by the following opposing rule-based discrete-time SFJ model,}
  \begin{equation}
    x_i(k+1)=\beta_ix_i(0)+(1-
    \beta_i)\sum_{j =1}^{n}q_{ij}x_j(k)
    \label{eq:opinion_dynamics}
\end{equation} 
where 
$\beta_i$ denotes the degree of stubbornness of agent $i$ towards its initial opinion. 
The matrix $Q=[q_{ij}]$ is defined as:
\begin{align}
\label{eq:weighted_Adjacency}
q_{ij}=\begin{cases}
    \frac{a_{ij}}{\sum_{j=1}^{n}|a_{ij}|} & \text{ if }\sum_{j=1}^{n}|a_{ij}| \neq 0  \\
   1 & \text{ if }\sum_{j=1}^{n}|a_{ij}| = 0 \text{ and } i=j\\
   0 & \text{ if }\sum_{j=1}^{n}|a_{ij}| = 0 \text{ and } i\neq j
\end{cases}    
\end{align}
 {We assume that the self-loop weight $a_{ii}>0$ for $i\in \{1,2,...,n\}$.}
An agent with $\beta_i>0$ is stubborn and the set of all the stubborn agents in $\mathcal{G}$ is denoted by $\mathcal{V}_{S}$.

\begin{remark}
\label{remark:stubbornness_of_single_OL}
A fully stubborn agent ($\beta_i=1$) and a sink in the network $\mathcal{G}$ are equivalent in the sense that an agent's opinion remains unchanged in either of the cases.  {In this work, we consider such agents to be sinks in $\mathcal{G}$.} An agent which is not a sink of $\mathcal{G}$ has $\beta_i<1$, so $\beta_i \in [0,1)$ for all $i\in \mathcal{V}$. \end{remark}

Finally, the SFJ model in vector form is given by,
\begin{equation}
\label{vector_op_model}
    \mathbf{x}(k+1)=(I-\beta)Q\mathbf{x}(k)+\beta \mathbf{x}(0)
\end{equation}
where 
the matrix $\beta=diag([\beta_1,\beta_2,...,\beta_n])$ is a diagonal matrix.
We are interested in the steady-state behaviours of the opinions arising under the  {opposing-rule based SFJ model}. To ease this analysis, we present the following classification of the agents. 
\subsection{Classification of agents}
\label{subsec:COA}
\begin{figure}[h]
\centering
  \begin{subfigure}{0.3\textwidth}
  \centering
    \includegraphics[width=1\linewidth]{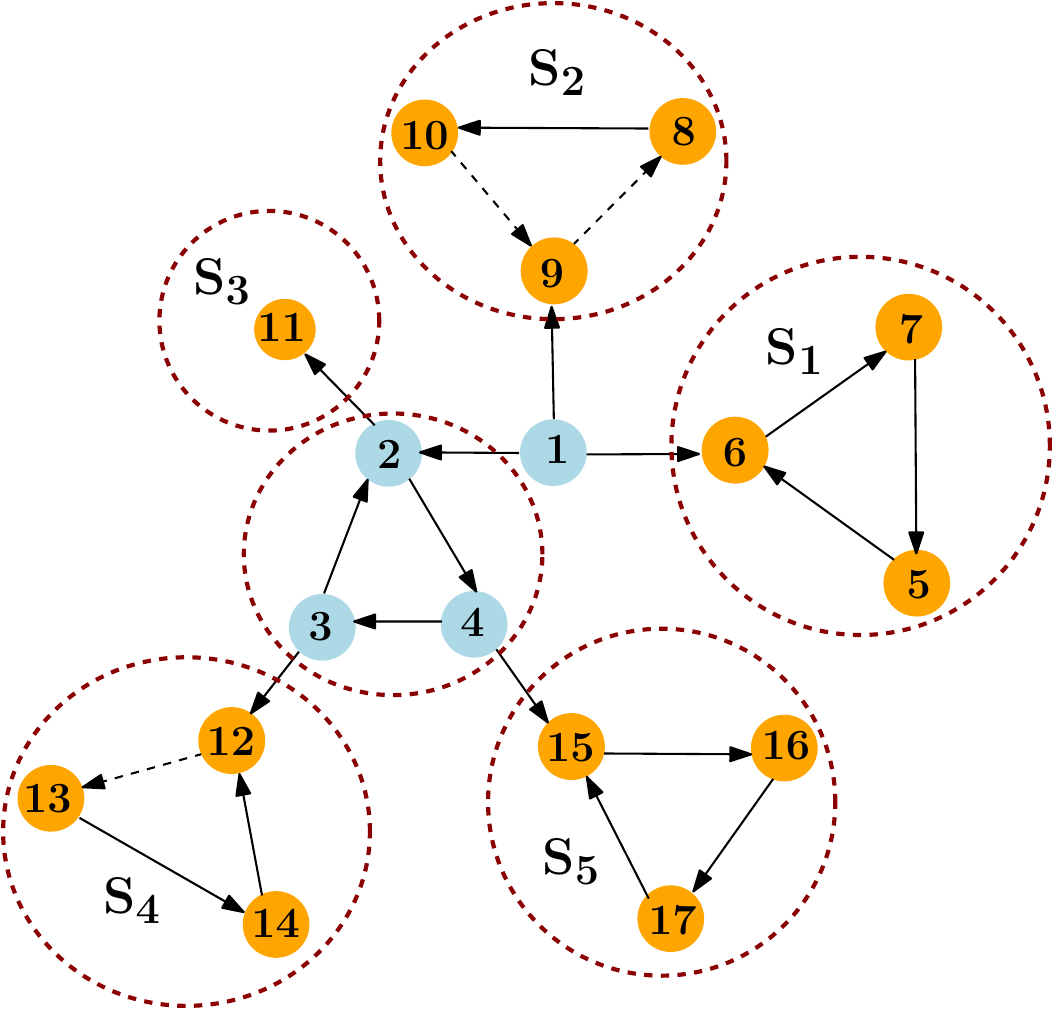}
    \caption{A weakly connected digraph \\ $\mathcal{G}$ with circled SCCs.}
    \label{fig_1_1}
  \end{subfigure}%
  \begin{subfigure}{0.2\textwidth}
  \centering
    \includegraphics[width=1\linewidth]{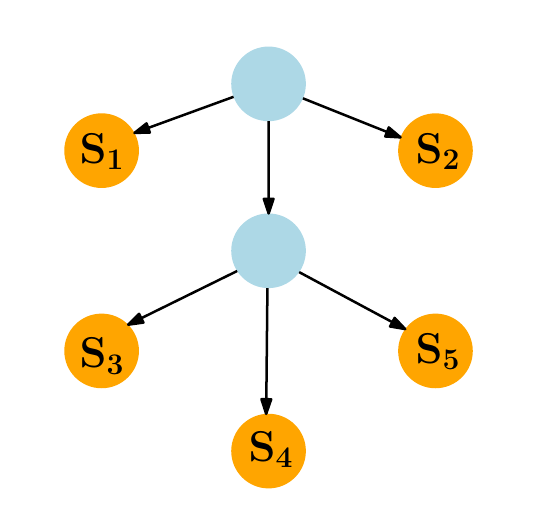}
    \caption{The condensation graph $C(\mathcal{G})$ of network $\mathcal{G}$}
     \label{fig_1_2}
  \end{subfigure}
  \caption{The followers are shown in blue and the OLs in orange. The solid and dotted edges denote cooperative and antagonistic interactions, respectively.}
\end{figure}

Social networks, in general, are weakly connected with strongly connected subgroups of individuals formed on the basis of shared interests, geographical locations, culture, \textit{etc.} Considering these strongly connected subgroups as nodes, the condensation graph $C(\mathcal{G})=(\mathcal{V}_c,\mathcal{E}_c)$ is derived from the network $\mathcal{G}$. For a weakly connected network $\mathcal{G}$, its $C(\mathcal{G})$ is a directed acyclic graph which comprises one or more sinks. We define $\mathcal{S}$ as the set comprising of the sinks of $C(\mathcal{G})$.
The number of sinks in graph $C(\mathcal{G})$ is denoted by $n_s=|\mathcal{S}|$. Note that a sink of $C(\mathcal{G})$ can be a set of nodes in the graph $\mathcal{G}$.
\begin{expm}
   Let us consider a network of agents represented by $\mathcal{G}=(\mathcal{V},\mathcal{E},A)$ in Fig. \ref{fig_1_1} and with its condensation graph $C(\mathcal{G})=(\mathcal{V}_c,\mathcal{E}_c)$ in Fig. \ref{fig_1_2}. The network $\mathcal{G}$ is weakly connected and has sinks $\mathcal{S}=\{S_1,S_2,..,S_5\}=\{ \{5,6,7\},$ $
\{8,9,10\},\{11\},\{12,13,14\},\{15,16,17\}\}$. 
\end{expm} 
   
This subsection focuses on arriving at a classification of the agents in a network regarding their ability to influence the other agents. We begin with the identification of the opinion leaders in the network.
 In a network $\mathcal{G}$ with opinions evolving according to eqn. \eqref{eq:opinion_dynamics}, an agent which belongs to a sink $S_i$ of $C(\mathcal{G})$ interacts only with the other agents in $S_i$. As a result, its opinion remains unaffected by the agents in $\mathcal{G}$ not belonging to $S_i$. On the other hand, a node in $\mathcal{G}$ not belonging to any sink of $C(\mathcal{G})$ must have an edge or a directed path to the node(s) belonging to a sink of $C(\mathcal{G})$. This occurs because $C(\mathcal{G})$ is a directed acyclic graph, Thus, the latter's opinion depends on the former during opinion evolution. This results in the following classification of agents:
   \begin{itemize}
    \item An opinion leader (henceforth, OL) is a node of network $\mathcal{G}$ belonging to a sink $S_i$ of  $C(\mathcal{G})$. The set of all the OLs is defined as $\mathcal{V}_{o}$. 
    \item  A follower in the network $\mathcal{G}$ does not belong to any sink of $C(\mathcal{G})$. The set of followers $\mathcal{V}_{F} \coloneqq \mathcal{V}\setminus \mathcal{V}_{o}$
\end{itemize}

In Fig. \ref{fig_1_1}, the OLs in $\mathcal{G}$ are $\mathcal{V}_o$ are $\{5,6,...,17\}$ and the rest are followers. 
\begin{remark}
 {Note that while the self-loop weights $a_{ii}$ can be non-negative for the followers, they must be strictly positive for OLs under the SFJ model \eqref{vector_op_model}. This assumption is justified as each OL, situated in topologically prominent locations in the graph, is naturally self-confident, which is modelled as $a_{ii}>0$.}
\end{remark}

Henceforth, we classify the sinks of $C(\mathcal{G})$ containing the OLs based on the nature of the interactions among OLs: 
\begin{itemize}
    \item A sink $S_i$ of $C(\mathcal{G})$ is called a \textit{SB sink} if the associated OLs in $S_i$ form a SB subgraph in $\mathcal{G}$ which can be partitioned into two groups based on signed interactions as defined in Sec. \ref{subsec:GP}. 
    \item  A sink $S_i$ of $C(\mathcal{G})$ is called \textit{SUB sink} if the associated OLs in $S_i$ form a SUB subgraph in $\mathcal{G}$.
\end{itemize}
Structural balance property implies that the relations among agents in a group satisfy Heider's Laws \cite{heider1946attitudes}. 
In general, agents with cooperative interactions are considered to be SB \cite{altafini}.  {Thus, a sink having only cooperative interactions amongst OLs is also a SB sink. Additionally, we consider a sink in $C(\mathcal{G})$, which is a single node that forms a sink in $\mathcal{G}$ as well, as a SB sink. In Fig. \ref{fig_1_1}, the sinks $S_1,S_2,S_3$ and $S_5$ of  $C(\mathcal{G})$ are SB sinks and $S_4$ is a SUB sink.}

Finally, given the presence of stubborn agents in the network, $\mathcal{V}_{S}$ comprises both stubborn followers and stubborn OLs.  {As discussed in Remark \ref{remark:stubbornness_of_single_OL}, the OL in a sink in $\mathcal{G}$ is considered non-stubborn.} Hence, stubbornness may arise only in a sink of $C(\mathcal{G})$ composed of two or more OLs. We define a set $\mathcal{S}_{ns}$ to distinguish the sinks in $C(\mathcal{G})$ which consist of non-stubborn OLs and form a SB subgraph as:
\vspace{-5pt}
\begin{align}
\label{eqn:S_n}
   \mathcal{S}_{ns} \coloneqq \{S_i\in \mathcal{S}: S_i & \text{ is SB}  \text{ and }  \nonumber \\ & S_i \subseteq \mathcal{V}\setminus \mathcal{V}_{S}\} 
\end{align}
The classification of agents discussed in this section allows us to analyse the effect of different kinds of OLs in the network.

\section{Convergence analysis}
    \label{Sec4}
In this section, we study the convergence of opinions of agents evolving according to eqn. \eqref{eq:opinion_dynamics} in a weakly connected signed digraph. The nodes of a weakly connected graph can be suitably permuted such that the adjacency matrix becomes block triangular. Therefore, we renumber the nodes such that $i=\{1,2,...,m\}$ are the followers and the rest are OLs, where the OLs associated with a sink of $C(\mathcal{G})$ are grouped together, resulting in a block triangular $A$. Henceforth, we will use this numbering of nodes throughout the paper. Further, we define the matrix $P$ as $P=(I-\beta)Q$. By definition, $P$ is block triangular of the following form,
\begin{align}
\label{B_rearranged}
P=\begin{bmatrix}
 P_{11} & P_{12} & ...   & P_{1(n_s+1)} \\
 \mathbb{0} &  P_{22} & \mathbb{0} & \mathbb{0}\\
  \vdots & ...  & \ddots & \vdots \\
 \mathbb{0} & \mathbb{0} & \mathbb{0} & P_{(n_s+1)(n_s+1)} \\
\end{bmatrix}
\end{align}
where $P_{ij}$ are submatrices for $i,j\in \{1,...,n_s+1\}$. The network is weakly connected; hence, each follower has a path to at least one of the OLs. Thus, $P_{1j} \neq \mathbb{0}~\forall \ j \in \{2,...,n_s+1\}.$
\begin{thm}
\label{thm:spectral_properties_P}
Consider a weakly connected signed digraph $\mathcal{G}$, the associated matrix $P$ is semi-convergent (and not convergent) if and only if $\mathcal{S}_{ns}$ (defined in \eqref{eqn:S_n}) is non-empty. Otherwise, $P$ is convergent.
\end{thm}
\begin{proof}
We know that $P$ is block triangular, hence, $\Spec(P)=\cup_{i=1}^{(n_s+1)}\Spec(P_{ii})$. So, we analyse the spectrum of each submatrix $P_{ii}$ for $i \in \{1,...,n_s+1\}$. Consider the submatrix $P_{11}$ associated with followers in the network. We define the matrix $\tilde{P}_{11}$ derived from $P_{11}$ as $\tilde{P}_{11}=[|p_{ij}|]$. In order to determine the spectral properties of $P_{11}$, we analyse the spectrum of the associated non-negative matrix $\tilde{P}_{11}$.

For follower $i$ in the network, the following scenarios may occur:
\begin{itemize}
    \item[(a)] If $i$ is not stubborn and all of its out-neighbours are followers, then the row-sum for follower $i$ in $\tilde{P}_{11}$ is equal $\sum_{j\in \mathcal{V}_{F}}|p_{ij}|=1$.
    \item[(b)] If the follower $i$ is stubborn or some of its neighbours are OLs, or both, then the row sum for follower $i$ in $\tilde{P}_{11}$ is $\sum_{j\in \mathcal{V}_{F}}|p_{ij}|<1$.
\end{itemize}
When none of the followers satisfy condition $(a)$, then it directly follows that $\rho(\Tilde{P}_{11})<1$ as $\rho(\Tilde{P}_{11})\leq \|\Tilde{P}_{11}\|_{\infty}<1$. If there exists one or more followers that satisfy condition $(a)$, then $\Tilde{P}_{11}$ is row substochastic. Since the network is weakly connected, each follower has a directed path to one or more OLs. Thus, every follower with a row-sum equal to $1$ in $\tilde{P}_{11}$ has a path to a follower $j$ whose neighbour is an OL. This implies that $\rho(\Tilde{P}_{ii})<1$ \cite{FB-LNS}. Consequently, from Lemma \ref{lm:2}, we infer that $\rho(P_{11})<1$.

Now, we discuss the spectral properties of the blocks $P_{ii}$ corresponding to the OLs in sink $S_{i-1} \in \mathcal{S}$ for $i\in \{2,3,...,n_s+1\}$. If none of the OLs in ${S}_{i-1}$ are stubborn, then the following scenarios can arise. 
        \begin{itemize}
            \item Suppose the sink $S_{i-1}$ is a SB sink. Then, the spectrum of $P_{ii}$ has a simple eigenvalue $1$ and the other eigenvalues have magnitudes strictly less than one \cite{xia2015structural}.
            
            \item Suppose the sink $S_{i-1}$ is a SUB sink. Then, each eigenvalue of $P_{ii}$ has magnitude strictly less than one \cite{xia2015structural}. 
        \end{itemize}

Let us consider the case when a sink $S_{i-1}$ is associated with one or more OLs who are stubborn. 
We know that if an OL $j \in S_{i-1}$ is not stubborn, then $\sum_{k \in S_{i-1}}|p_{jk}|=1$, and $\sum_{k \in S_{i-1}}|p_{jk}|<1$ for a stubborn $j$. Thus, the matrix $\tilde{P}_{ii}$ is row substochastic. Since the OLs in sink $S_{i-1}$ are strongly connected, the spectral radius of $\rho(\tilde{P}_{ii})<1$ \cite{FB-LNS}. We deduce from Lemma \ref{lm:2} that $\rho(P_{ii})<1$ for each $i\in\{2,...,n_s+1\}$. 

The preceding discussions imply that $P$ is semi-convergent (and not convergent) if and only if the network $\mathcal{G}$ possesses an OL(s) belonging to a SB sink in $C(\mathcal{G})$ such that none of them is stubborn.  Thus, it suffices to have a non-empty $\mathcal{S}_{ns}$. Otherwise, $\rho(P)<1$ and it is convergent.
\end{proof}

Theorem \ref{thm:spectral_properties_P} relates the spectral properties of $P$ with the topological properties of $\mathcal{G}$ and stubborn behaviour. 
Based on the spectral properties of $P$, we present the steady-state behaviours under the SFJ model \eqref{vector_op_model} in the following results. 
\begin{corollary}
\label{Cor:S_n_phi}
\label{thm:spectral_properties_P}
 {Consider a digraph $\mathcal{G}$ with opinions evolving under the proposed SFJ model \eqref{vector_op_model}}. If $\mathcal{S}_{ns}$ is empty, the final opinions converge to, \begin{align}
\label{eqn:z_s_Sn_0}
 {\mathbf{x}^*}=(I-P)^{-1}\beta \mathbf{x}(0)   
\end{align}
where ${\mathbf{x}^*}=[x_1^*,...,x_n^*]=\lim_{k \to \infty}\mathbf{x}(k)$.
\end{corollary}
\begin{proof}
At steady state, the opinions of agents under \eqref{vector_op_model} satisfy: $ \mathbf{x}^*=P\mathbf{x}^*+\beta \mathbf{x}(0)$. From Theorem \ref{thm:spectral_properties_P} it follows that  if $\mathcal{S}_{ns}=\emptyset$, then $\rho(P)<1$. Thus, $(I-P)$ is invertible and the final opinion converges to \eqref{eqn:z_s_Sn_0}.
\end{proof}

In eqn. \eqref{eqn:z_s_Sn_0}, the term $\beta \mathbf{x}(0)$ results in a vector with zero entries pertaining to the non-stubborn agents. Thus, when $\mathcal{S}_{ns}$ is empty, the final opinion of every agent in the network depends only on the initial opinions of the stubborn agents. 

However, if  a digraph $\mathcal{G}$ has OLs belonging to a sink $S_i \in \mathcal{S}_{ns}$, Theorem \ref{thm:spectral_properties_P} demonstrates that $P$ is semi-convergent and not convergent. Meaning that  $\rho(P)=1$ and $(I-P)$ is not invertible. We analyse the overall steady-state behaviours in the presence of such OLs by decoupling the dynamics of followers and the OLs. We achieve this by
conformally partitioning $\beta$ as $P$ such that $\beta=diag(\beta_{11},...,$ $\beta_{(n_s+1)(n_s+1}))$ where $\beta_{11}$ gives stubbornness of the followers and $\beta_{ii}$ gives the stubbornness of OLs in sink $S_{i-1}$ of $C(\mathcal{G})$ for $i\in \{2,...,n_s+1\}$. Let the opinion vector $\mathbf{x}$ is also analogously partitioned as $\mathbf{x}=[\mathbf{x}_1,\mathbf{x}_{2},...,\mathbf{x}_{n_s+1}]$. Thus, opinions of followers governed by the opposing-rule based SFJ model \eqref{vector_op_model}, evolve as:
\begin{align}
\label{eqn:op_followers}
\mathbf{x}_1(k+1)&=P_{11}\mathbf{x}_1(k)+P_{12}\mathbf{x}_2(k)+\cdots+ \nonumber \\
    &P_{1(n_s+1)}\mathbf{x}_{(n_s+1)}(k)+\beta_{11}\mathbf{x}_1(0),
    \end{align}
while the opinions of OLs evolve as:
    \begin{align}  
    \label{eqn:op_leaders}
    \mathbf{x}_i(k+1)=P_{ii}\mathbf{x}_i(k)+\beta_{ii}\mathbf{x}_i(0) \qquad \textit{if } &S_{i-1} \notin \mathcal{S}_{ns}  \nonumber \\
\mathbf{x}_i(k+1)=P_{ii}\mathbf{x}_i(k), \qquad \textit{if } &S_{i-1} \in \mathcal{S}_{ns}
\end{align}
The following result presents the steady-state behaviours under this case.
\begin{corollary}
\label{corollary:non_empty}
   {Consider a digraph $\mathcal{G}$ with opinions evolving under the proposed SFJ model \eqref{vector_op_model}}. If $\mathcal{S}_{ns}$ is not empty, the final opinions converge as follows, \begin{align}
\label{eqn:z_s_followers}
 \mathbf{x}_1^*&=(I-(I-\beta_{11})P_{11})^{-1}\big(\sum_{j\in \mathcal{S}_{ns}}\mathbf{v}_i\mathbf{w}_i^T\mathbf{x}_i(0)+ \nonumber \\
 &\sum_{j \notin \mathcal{S}_{ns}}(I-(I-\beta_{ii})P_{ii})^{-1}\beta_{ii}\mathbf{x}_i(0)\big).&  \nonumber \\
 \mathbf{x}_i^*&=(I-(I-\beta_{ii})P_{ii})^{-1}\beta_{ii}\mathbf{x}_i(0), \quad S_{i-1}\notin \mathcal{S}_{ns} \nonumber \\\mathbf{x}_i^*&=\mathbf{v}_i\mathbf{w}_i^T\mathbf{x}_i(0), \quad S_{i-1}\in \mathcal{S}_{ns}
\end{align}   
where  $\mathbf{w}_i$ and $\mathbf{v}_i$ are the left and right eigenvectors, respectively, of $P_{ii}$ corresponding to the eigenvalue $1$, such that $\mathbf{v}_{i}^T\mathbf{w}_i=1$ and $i\in\{2,...,n_s+1\}$.
\end{corollary}
\begin{table*}[h!]
\centering
\begin{tabular}{|m{1.75cm}|m{3.0cm}|m{3cm}|m{4.5cm}|m{3.5cm}|}
\hline \centering \textbf{Network topology} &  \multicolumn{2}{|m{6.5cm}|}{\textbf{Special structures within the network $\mathcal{G}$}}  &  \centering \textbf{Final opinions}  &    
    \textbf{ \centering Influential nodes}
  \\
\hline
&     \multicolumn{2}{|m{6.5cm}|}{$\mathcal{G}$ does not have any stubborn agents and is SUB ($\mathcal{S}_{ns} = \emptyset$). } & The opinions converge to the neutral opinion $\mathbf{x}^*=0$.  & None of the OLs are influential.  \\
\cline{2-5}
\multirow{2}{=}{$\mathcal{G}$ is strongly connected. ($C(\mathcal{G})$ has a single sink).} &  \multicolumn{2}{|m{6.5cm}|}{  There exists atleast one stubborn agent in $\mathcal{G}$ ($\mathcal{S}_{ns} = \emptyset$).}    &  {The opinions converge to a non-trivial opinion given by \eqref{eqn:z_s_Sn_0}, often leading to disagreement}. &    Only the stubborn OLs are influential. \\ 
\cline{2-5}
& \multicolumn{2}{|m{6.5cm}|}{ $\mathcal{G}$ does not have any stubborn agents and has only cooperative interactions ($\mathcal{S}_{ns} \neq \emptyset$).}   & {The agents achieve consensus with final opinion given by \eqref{eqn:z_s_followers}.}        & All the OLs are influential. \\
\cline{2-5}
&   \multicolumn{2}{|m{6.5cm}|}{ $\mathcal{G}$ does not have any stubborn agents and is SB with atleast one antagonistic interaction ($\mathcal{S}_{ns} \neq \emptyset$).}                &  The agents achieve biparite consensus leading to polarised opinions given by \eqref{eqn:z_s_followers}.   & All the OLs are influential. 
\\
\hline

&  \multirow{2}{=}{Each sink in $C(\mathcal{G})$ is SUB ($\mathcal{S}_{ns} = \emptyset$).} &  There are no stubborn OLs. & {The opinions converge to the neutral opinion $\mathbf{x}^*=0$.}  & None of the OLs are influential.  \\
\cline{3-5}
&  & There exists atleast one sink in $C(\mathcal{G})$ with atleast one stubborn OL. & {The opinions converge to a non-trivial opinion given by \eqref{eqn:z_s_Sn_0}.} & Only the stubborn OLs are influential.  \\

\cline{2-5}
\multirow{4}{=}{$\mathcal{G}$ is weakly connected  ($C(\mathcal{G})$ can have multiple sinks)}
& \multirow{4}{=}{Each sink in $C(\mathcal{G})$ is SB and does not contain any stubborn OLs ($\mathcal{S}_{ns} \neq \emptyset$).} & 
  There exists a sink $S_i$ which has only cooperative interactions. & { The agents constituting $S_i$ achieve consensus. However, the overall behaviour of the group can be disagreement (given by \eqref{eqn:z_s_followers}) depending on the nature of the other sinks}. & Each OL in $S_i$ is influential; there maybe other influential agents in the group. \\
\cline{3-5} 
& & There exists a sink $S_i$ which has atleast one antagonistic interaction. & { The agents constituting $S_i$ achieve bipartite consensus. However, the overall behaviour of the group might not be polarised (given by \eqref{eqn:z_s_followers}) depending on the nature of the other sinks.} & Each OL in $S_i$ is influential; there may be other influential agents in the group.  \\

\cline{2-5}
& \multicolumn{2}{|m{6.5cm}|}{ There exists atleast one sink in $C(\mathcal{G})$ which is SB and contains atleast one stubborn OL.} &  {The opinions converge to a non-trivial opinion given by \eqref{eqn:z_s_Sn_0} or \eqref{eqn:z_s_followers}, often leading to disagreement.} & All the stubborn OLs and the non-stubborn OLs belonging to SBs without any stubborn OLs are influential.\\
\hline
\end{tabular}
\caption{The table shows the emergent collective behaviours for varying network topologies under the proposed SFJ model \eqref{vector_op_model}. It highlights the fact that the presence of stubborn agents within SB subgraphs of $\mathcal{G}$ always leads to disagreement within the agents, irrespective of the nature of every other sink.}
\label{table:emergent}
\end{table*}
\begin{proof}
The opinions of OLs within a sink evolve independently of other agents in the network, as shown in eqn. \eqref{eqn:op_leaders}. Thus, we begin by determining their final opinions:
\begin{itemize}
    \item If the OLs belong to a sink $S_{i-1} \notin \mathcal{S}_{ns}$, we know by Theorem \ref{thm:spectral_properties_P} that $\rho(P_{ii})<1$. As a result, their final opinions converge to: $\mathbf{x}_i^*=(I-(I-\beta_{ii})P_{ii})^{-1}\beta_{ii}\mathbf{x}_i(0)$.
    \item However, if the OLs belong to sink $S_{i-1} \in \mathcal{S}_{ns}$, then by Theorem \ref{thm:spectral_properties_P}, $\rho(P_{ii})=1$ is simple and strictly larger than the magnitude of the rest of the eigenvalues. Thus, the final opinions of OLs converge to $\mathbf{x}_i^*=\mathbf{v}_i\mathbf{w}_i^T\mathbf{x}_i(0)$. \cite{altafini}
  \end{itemize} 
Next, we determine the final opinions of followers, which depend on the opinions of both stubborn followers and OLs, as shown in the eqn. \eqref{eqn:op_followers}.  We know from Theorem \ref{thm:spectral_properties_P} that the $\rho(P_{11})<1$. Thus, at steady state, we can obtain the final opinions of the followers from eqn. \eqref{eqn:op_followers} by substituting the final opinions of the OLs and inverting $(I-P_{11})$, which results in eqn. \eqref{eqn:z_s_followers} 
\end{proof}

Corollary \ref{corollary:non_empty} presents the final opinions of the agents in $\mathcal{G}$ under the proposed SFJ model \eqref{vector_op_model}. It highlights that if 
a sink of $C(\mathcal{G})$ has a stubborn OL(s), only such OLs affect the final opinions. 
In sinks without the stubborn agents, the nature of interactions determines the final outcome. For instance, the OLs in a SUB sink that are unaffected by stubbornness converge to $0$ and do not influence any agent. Again, based on the nature of interactions, the OLs in a SB sink can have the following variety of behaviours,
    \begin{itemize}
        \item If the OLs in $S_{i-1}$ have cooperative interactions amongst them, then consensus occurs among the OLs. 
        \item If the OLs in $S_{i-1}$ have antagonistic interactions, then bipartite consensus occurs where the opinions are split in opposing views based on the bipartition in $S_{i-1}$ due to structural balance.
        \item A special scenario occurs when $S_{i-1}$ consists of a single node forming a sink in $\mathcal{G}$. In this case, the opinion of this OL does not change.
    \end{itemize}
It is simple to show that for each $S_{i-1} \in \mathcal{S}_{ns}$, the entries of $\mathbf{w}_i$ are non-zero. Consequently, the initial opinion of each OL in $S_{i-1}$ affects the opinions of the followers and the OLs in $S_{i-1}$.
From the above discussion, it becomes clear that stubborn agents and OLs who do not interact with any other stubborn agents are capable of influencing the opinions of the followers.

\begin{remark}
\label{rem:4}
Corollary \ref{corollary:non_empty} shows that if we introduce stubbornness in an OL in a sink, the non-stubborn OLs within that sink lose their influence. This scenario mirrors the impact of  \textit{veto power} held by the permanent members of the United Nations Security Council \cite{US_hegemony}. Once the \textit{stubbornness} (or \textit{veto power}) is enforced, the opinions of the other members of the group become inconsequential.  
\end{remark} 

\begin{remark}
\label{remark:comparison_signed}
In \cite{razaq2025signed}, the authors present 
sufficient conditions that ensure convergence under the repelling rule-based SFJ model, which includes that the matrix $Q$ must be Eventually Stochastic (ES). By definition, $Q$ is ES if there is a $k_0 \in \mathbb{N}$ such that $Q^k>0$ for all $k\geq k$ and $Q \mathbb{1}=\mathbb{1}$. Note that unless the network is strongly connected, $Q^k$ cannot be positive. Thus, the opinions under the repelling SFJ might not converge for certain weakly connected signed digraphs. On the contrary, Theorem \ref{thm:spectral_properties_P} shows that under the opposing-rule based SFJ model \eqref{vector_op_model}, the opinions converge for any weakly connected digraphs.
\end{remark}
In this section, we discussed the role of the nature of interactions and stubbornness in deciding the influential agents in the network. Next, we quantify the influence each of these agents exert over the others in the network.

\section{Absolute Influence centrality}
\label{Sec6}
In this subsection, we propose a centrality measure to determine the overall contribution of an influential agent in the final opinions of agents in the network. We use derivations of final opinions $\mathbf{x}^*$ to construct a matrix $\Theta=[\theta_{ij}]$ such that,
\begin{align}
\label{eq:inf_x0}
    \mathbf{x}^*=\Theta\mathbf{x}(0)
\end{align}
The entries of $\Theta$ account for the effect of the initial opinions of the influential agents on the final opinion vector $\mathbf{x}^*$.
The antagonism in the network can cause the influence of an agent to be positive or negative. So, we characterise the influence using $\tilde\Theta=[|\theta_{ij}|]$ to determine the exact contribution of an agent in $\mathbf{x}^*$. 
\begin{definition}
The absolute influence centrality vector quantifies the exact impact of the initial opinion of each agent in the final opinion pattern under the opposing-rule based SFJ model. Mathematically, it is defined as:
\begin{align}
\label{eqn:abs_influence_cent}
\mathbf{c}=\mathbf{\tilde{\Theta}}^T \mathbb{1}    
\end{align}  
where $\mathbf{c}$ is the absolute influence centrality vector.
\end{definition}
The above formulation allows us to determine the most influential agent which is simply the one with the highest \textit{absolute influence centrality}. As opposed to IC, the proposed centrality measure is applicable for signed networks and accounts for the influence of non-stubborn agents as well. 
\section{Simulation Results}
\label{sec:Simulation_Results}
In this section, we demonstrate our simulation results on the Bitcoin Alpha dataset, which is a signed digraph comprising $3,783$ nodes and $24,186$ edges, of which $1,536$ have negative signs. A node in this network represents a Bitcoin trader, and the directed edges represent trust or distrust with weights ranging from $-10$ to $10$ based on ratings assigned by each trader to others. 

\begin{figure}[h]
    \centering
    \includegraphics[width=0.9\linewidth]{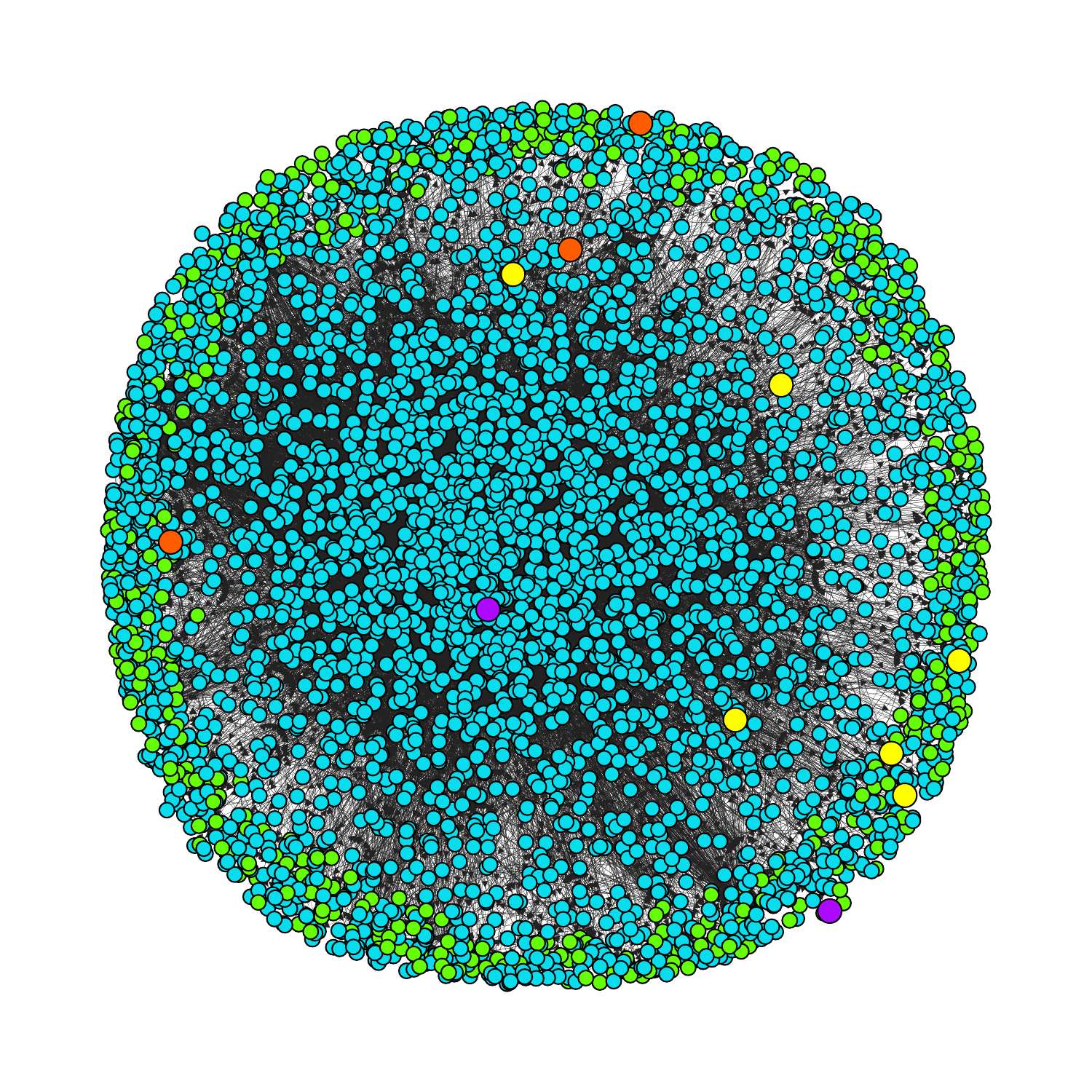}
    \caption{This is a visual representation of Bitcoin Alpha dataset. All stubborn agents are highlighted in yellow. The non-stubborn OLs in SB and SUB sinks are highlighted in orange and purple, respectively. Rest of the OLs are coloured green. The non-stubborn followers are shown in blue and the black lines represent the interactions between the agents.}
    
    \label{fig:Bitcoin_Alpha_dataset}
\end{figure}
\begin{figure}[h]
    \centering
    \includegraphics[width=1\linewidth]{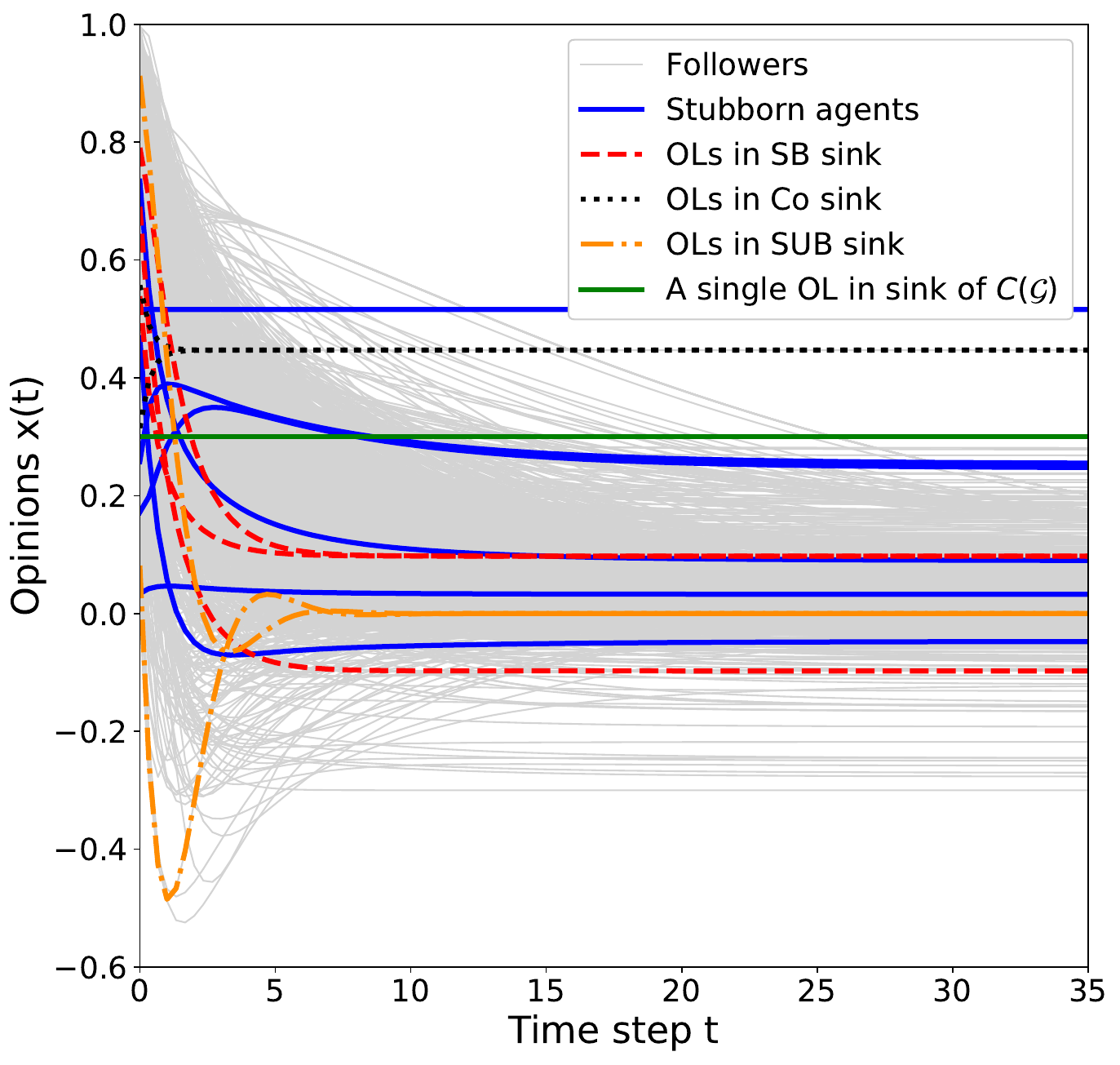}
    \caption{Opinion evolution under the proposed SFJ model.}
    \label{fig:OPINION_PLOT_Bitcoin_Alpha_dataset}
\end{figure}
\begin{figure}[h]
    \centering
    \includegraphics[width=1\linewidth]{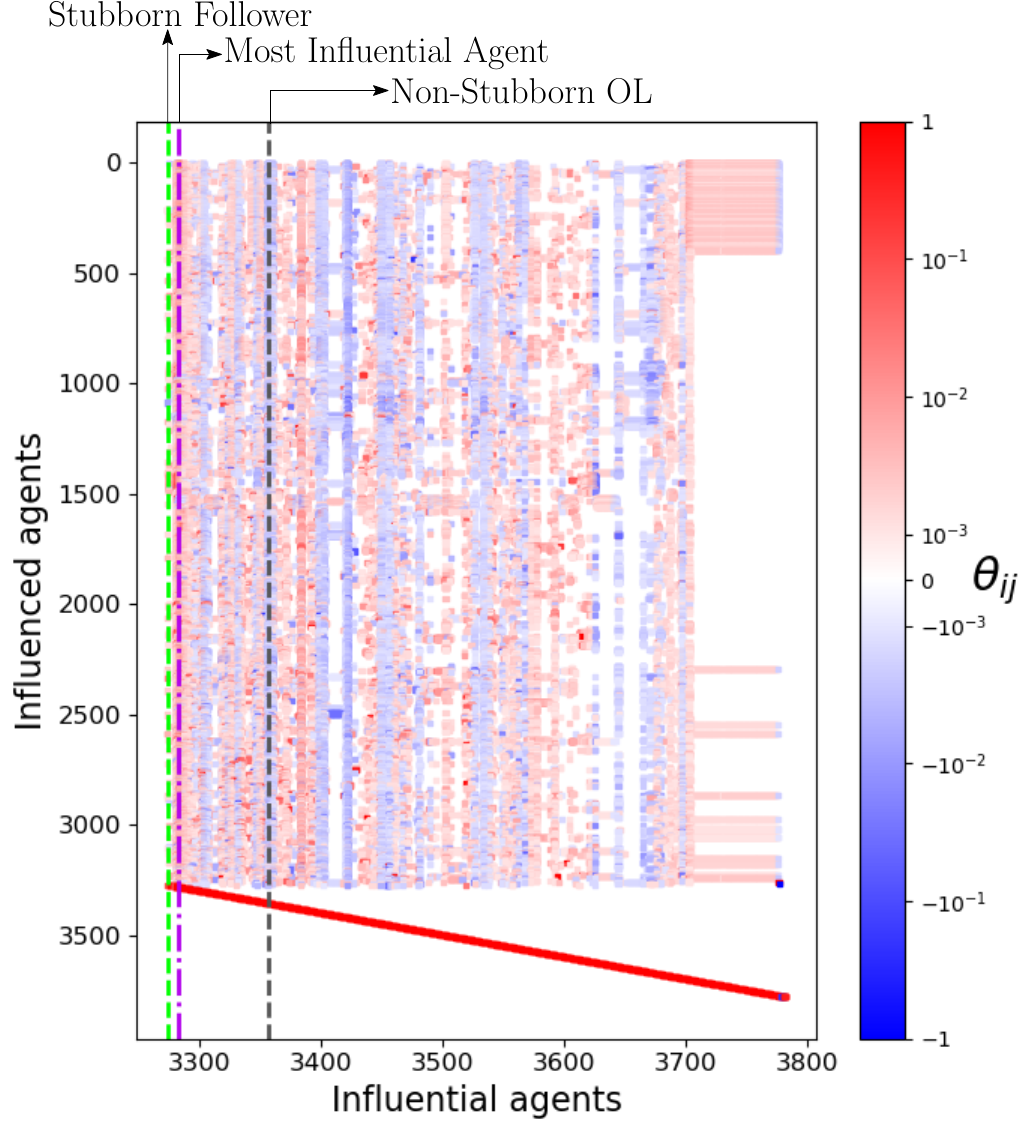}
    \caption{This is a scatter plot representation of the influence matrix $\Theta$ for the Bitcoin Alpha data set. The abscissa shows the agents which are \textit{influential}. The ordinate includes the agents which get influenced. Using the proposed absolute centrality measure, we determine the central node (indexed as 1163 in the dataset and as $3282$ in the paper) in the network; its influence over others is indicated by the dashed vertical line in purple. The vertical dashed line in green represents the influence of a stubborn follower (indexed as 1044 in the dataset and as $3274$ in the paper), which is more influential than several other OLs (for example, the OL indexed as $3244$ in the dataset and as $3357$ in the paper).}
    \label{fig:scatter_plot}
\end{figure}
The condensation graph $C(\mathcal{G})$ obtained for this network has $502$ sinks, out of which $497$ have single OLs and the remaining $5$ have multiple OLs with only cooperative interactions. We modify the signs of three edges to convert two out of these multi-OL sinks into a SB and a SUB sink, respectively. Subsequently, we choose the stubborn agents such that the OLs belonging to two of the remaining cooperative sinks in $C(\mathcal{G})$ become stubborn. Consequently, out of $5$ sinks with multiple OLs: two have stubborn agents, one is SB, one is SUB and one has cooperative interactions. Thus, $\mathcal{S}_{ns} \neq \emptyset$ and we have each of the discussed diverse categories of sinks of $C(\mathcal{G})$. Additionally, we choose the $3$ followers to be stubborn to make them influential. This modified graph $\mathcal{G}$ with chosen stubborn agents is illustrated in Fig. \ref{fig:Bitcoin_Alpha_dataset}. 

We arrange the nodes in the network such that the agents that are never influential are indexed first, followed by those who can be influential. Therefore, (a) the first 3271 nodes are non-stubborn followers, (b) the nodes numbered 3271 to 3273 are non-stubborn OLs in SUB sinks (c) the nodes numbered 3274 to 3277 are stubborn followers, (d) the rest of the nodes are remaining OLs. We again highlight that the nodes corresponding to a sink in $C(\mathcal{G})$ are numbered sequentially. This leads to a different indexing compared to that used for the original dataset. For the ease of reference, we mentioned the original index followed by this alternate index in parentheses. 

In this example, we consider a scenario when the opinions of the agents evolve under the proposed SFJ model \eqref{vector_op_model}, with the stubbornness values and the initial opinions chosen from the uniform distribution over $[0,1]$. The evolution of the opinions over time is shown in Fig. \ref{fig:OPINION_PLOT_Bitcoin_Alpha_dataset}. As predicted by the proposed results: 
\begin{itemize}
    \item the OLs in cooperative sink (referred to as a Co sink in Fig. \ref{fig:OPINION_PLOT_Bitcoin_Alpha_dataset}) in $C(\mathcal{G})$ have consensus (black dotted line),
    \item the OLs in SUB sinks reach the neutral opinion (orange dash-dot line),
    \item the OLs in SB sink have bipartite consensus (red dashed lines), and,
    \item the opinions of those in single sinks remain at their initial values (green line).
\end{itemize}
 The opinions of stubborn agents and the followers are denoted in blue and grey, respectively. Note that, while the initial opinions lie in the range $[0,1]$, the final opinions lie outside of the convex hull of the initial opinions. This is a consequence of the impact of the negative influence some agents have over others.

We illustrate the impact of only the influential agents over all the agents in the network by using the scatter plot of the matrix $\Theta$ shown in Fig. \ref{fig:scatter_plot}. The positive and negative entries $\theta_{ij}$ are highlighted in red and blue. Just to recall, the entry $\theta_{ij}$ denotes the positive or negative influence of an influential agent $j$ on the influenced agent $i$. Through this plot, it is evident that both stubborn followers and OLs, affect the final opinions of the group. Finally, we identify the most influential agent using the proposed \textit{absolute influence centrality} measure given in eqn. \eqref{eqn:abs_influence_cent}. It reveals that the OL indexed $1163$ $(3282)$, which belongs to a single sink in $C(\mathcal{G})$, is the most influential agent despite being non-stubborn. Interestingly, there exists a (stubborn) follower $1044$ ($3274$) which is more influential than several (non-stubborn) OLs. These results highlight the importance of both topology and stubbornness in making an agent influential.

\section{Conclusions}
\label{sec:con}
Real-world social networks are often signed in nature and consist of stubborn individuals. The paper proposes an opposing rule-based SFJ model \eqref{vector_op_model} which explores the complex interplay of signed interactions, network topology and stubbornness in the opinion formation process. We begin the analysis by classifying the agents as: (a) OLs, which are topologically prominent agents in the network (b) followers, constituted by the rest of the agents. Further, we allow any agent in the network to be stubborn. In the given framework, a key advantage of the proposed model is that the final opinions always converge (to the behaviours of consensus, polarisation or disagreement) under all circumstances (arbitrary network topologies, stubbornness values and initial conditions), unlike in \cite{razaq2025signed}. The diverse range of emergent behaviours depends on the signed interactions, network connectivity and the position of stubborn agents and has been summarised conclusively in Table \ref{table:emergent}.

In the signed-Degroot framework, the OLs in SB and SUB sinks converge to polarised (bipartite consensus) and neutral opinions, respectively. On the contrary, we show that the introduction of stubbornness results in disagreement in both of these cases. Further, we prove that the emergent behaviours of the overall network are dictated either by the OLs (Corollary \ref{corollary:non_empty}) or the stubborn individuals (Corollaries \ref{Cor:S_n_phi} and \ref{corollary:non_empty}) or both. Yet another important observation is that not all of the OLs are \textit{influential} even though they have a topological advantage; this behaviour arises for non-stubborn OLs in two scenarios: (a) they are present in SUB sinks or (b) they have directed paths (or interactions) to stubborn OLs. Furthermore, signed interactions result in negative influences; due to this, the opinions of certain agents can even go outside of the convex hull of initial opinions. The numerical simulation presented in Sec. \ref{sec:Simulation_Results} highlights the same (please see Fig. \ref{fig:OPINION_PLOT_Bitcoin_Alpha_dataset}). 



Yet another contribution of the paper is that we propose the \textit{absolute influence centrality} measure. This centrality measure allows us to determine the overall influence of all the agents and also identify the most \textit{influential} ones in any arbitrary signed network. To the best of our knowledge, the proposed centrality measure is the first that accounts for the impact of stubborn behaviour and OLs simultaneously, that too in the presence of signed interactions. We demonstrate the applicability of the proposed measure even for a large data set (Bitcoin Alpha dataset, which contains $3783$ nodes) in Sec.\ref{sec:Simulation_Results}. While we can identify the \textit{influential} agents using the proposed centrality measure, we do not focus on modifying their influences in this paper. However, in practical applications like brand marketing, targeted awareness campaigning, electioneering, \textit{etc.}, it becomes important for some agents to be more influential than others. In future, we plan to analyse the approaches (suitable topological or stubbornness modifications in a network) that can alter the influence of certain chosen agents in a desired manner.
\bibliographystyle{elsarticle-num}
\bibliography{references_2}

\begin{thebibliography}{10}
\expandafter\ifx\csname url\endcsname\relax
  \def\url#1{\texttt{#1}}\fi
\expandafter\ifx\csname urlprefix\endcsname\relax\def\urlprefix{URL }\fi
\expandafter\ifx\csname href\endcsname\relax
  \def\href#1#2{#2} \def\path#1{#1}\fi

\bibitem{voramontri2019impact}
D.~Voramontri, L.~Klieb, Impact of social media on consumer behaviour, International Journal of Information and Decision Sciences 11~(3) (2019) 209--233.

\bibitem{fernandez2014voter}
J.~Fern{\'a}ndez-Gracia, K.~Suchecki, J.~J. Ramasco, M.~San~Miguel, V.~M. Egu{\'\i}luz, Is the voter model a model for voters?, Physical Review Letters 112~(15) (2014) 158701.

\bibitem{Gorodnichenko2021}
Y.~Gorodnichenko, T.~Pham, O.~Talavera, Social media, sentiment and public opinions: Evidence from {B}rexit and {U.S.} election 136 (7 2021).

\bibitem{degroot1974reaching}
M.~H. DeGroot, Reaching a consensus, Journal of the American Statistical Association 69~(345) (1974) 118--121.

\bibitem{FRIEDKIN1997209}
N.~E. Friedkin, E.~C. Jonsen, Social positions in influence networks, Social Networks 19~(3) (1997) 209--222.

\bibitem{rainer2002opinion}
H.~Rainer, U.~Krause, Opinion dynamics and bounded confidence: Models, analysis and simulation, Journal of Artificial Societies and Social Simulation 5~(3) (2002).

\bibitem{Community_Cleavage}
N.~E. Friedkin, The problem of social control and coordination of complex systems in sociology: A look at the community cleavage problem, IEEE Control Systems Magazine 35~(3) (2015) 40--51.

\bibitem{GHADERI20143209}
J.~Ghaderi, R.~Srikant, Opinion dynamics in social networks with stubborn agents: Equilibrium and convergence rate, Automatica 50~(12) (2014) 3209--3215.

\bibitem{gionis2013opinion}
A.~Gionis, E.~Terzi, P.~Tsaparas, Opinion maximization in social networks, in: Proceedings of the 2013 SIAM international conference on data mining, SIAM, 2013, pp. 387--395.

\bibitem{parsegov2016novel}
S.~E. Parsegov, A.~V. Proskurnikov, R.~Tempo, N.~E. Friedkin, Novel multidimensional models of opinion dynamics in social networks, IEEE Transactions on Automatic Control 62~(5) (2016) 2270--2285.

\bibitem{TIAN2018213}
Y.~Tian, L.~Wang, Opinion dynamics in social networks with stubborn agents: An issue-based perspective, Automatica 96 (2018) 213--223.

\bibitem{FJ_Model}
N.~E. Friedkin, E.~C. Johnsen, Social influence and opinions, The Journal of Mathematical Sociology 15~(3-4) (1990) 193--206.

\bibitem{xia2015structural}
W.~Xia, M.~Cao, K.~H. Johansson, Structural balance and opinion separation in trust--mistrust social networks, IEEE Transactions on Control of Network Systems 3~(1) (2015) 46--56.

\bibitem{altafini}
C.~Altafini, Consensus problems on networks with antagonistic interactions, IEEE Transactions on Automatic Control 58~(4) (2013) 935--946.

\bibitem{fontan2022multiagent}
A.~Fontan, L.~Wang, Y.~Hong, G.~Shi, C.~Altafini, Multiagent consensus over time-invariant and time-varying signed digraphs via eventual positivity, IEEE Transactions on Automatic Control 68~(9) (2022) 5429--5444.

\bibitem{priya2024desired}
S.~Priya, A.~Shrinate, T.~Tripathy, Desired group consensus in arbitrary signed networks, Authorea Preprints (2024).

\bibitem{shrinate2023desired}
A.~Shrinate, T.~Tripathy, L.~Behera, Desired clustering in signed networks, in: 2023 Ninth Indian Control Conference (ICC), IEEE, 2023, pp. 233--238.

\bibitem{razaq2025signed}
M.~A. Razaq, C.~Altafini, Signed friedkin-johnsen models: Opinion dynamics with stubbornness and antagonism, IEEE Transactions on Automatic Control (2025).

\bibitem{Tian2022}
Y.~Tian, P.~Jia, A.~MirTabatabaei, L.~Wang, N.~E. Friedkin, F.~Bullo, Social power evolution in influence networks with stubborn individuals, IEEE Transactions on Automatic Control 67~(2) (2022) 574--588.

\bibitem{gelfand1941normierte}
I.~Gelfand, Normierte ringe 9~(1) (1941) 3--24.

\bibitem{ding2023electoral}
C.~Ding, W.~Jabr, H.~Guo, Electoral competition in the age of social media: The role of social media influencers., MIS Quarterly 47~(4) (2023).

\bibitem{NEAL2020103}
Z.~P. Neal, A sign of the times? {W}eak and strong polarization in the {U.S.} congress, 1973–2016, Social Networks 60 (2020) 103--112.

\bibitem{harary1961structural}
F.~Harary, A structural analysis of the situation in the {M}iddle {E}ast in 1956, Journal of Conflict Resolution 5~(2) (1961) 167--178.

\bibitem{fink2009off}
J.~S. Fink, H.~M. Parker, M.~Brett, J.~Higgins, Off-field behavior of athletes and team identification: Using social identity theory and balance theory to explain fan reactions, Journal of Sport Management 23~(2) (2009) 142--155.

\bibitem{heider1946attitudes}
F.~Heider, Attitudes and cognitive organization, The Journal of psychology 21~(1) (1946) 107--112.

\bibitem{FB-LNS}
F.~Bullo, Lectures on Network Systems, {1.6} Edition, Kindle Direct Publishing, 2022.

\bibitem{US_hegemony}
D.~J. Puchala, {World Hegemony and the United Nations}, International Studies Review 7~(4) (2005) 571--584.

\end{thebibliography}
\end{document}